\documentclass[prl,aps,twocolumn,amssymb]{revtex4}
\usepackage{amssymb}

\usepackage{graphicx}
\usepackage{amsmath}
\usepackage{times}
\usepackage{subfigure}

\begin{document}

\title{Self-trapping in an array of coupled 1D Bose gases}
\author{Aaron Reinhard, Jean-F\'{e}lix Riou, Laura A. Zundel and David S. Weiss}

\affiliation{ Physics Department,  The Pennsylvania State University, 104 Davey Lab, University Park, PA 16802}

\author{Shuming Li and  Ana Maria Rey}

\affiliation{JILA, NIST, Department of Physics, University of Colorado, 440, UCB, Boulder, CO, 80309}

\author{Rafael Hipolito}

\affiliation{Department of Engineering Science and Physics, City University of New York, College of Staten Island, Staten Island, NY 10314}

\date{\today}
\begin{abstract}
We study the transverse expansion of arrays of ultracold $^{87}$Rb atoms weakly confined in tubes created by a 2D optical lattice, and observe that transverse expansion is delayed because of mutual atom interactions. A mean-field model of a coupled array shows that atoms become localized within a roughly square fort-like self-trapping barrier with time-evolving edges. But the observed dynamics is poorly described by the mean-field model. Theoretical introduction of random phase fluctuations among tubes improves the agreement with experiment, but does not correctly predict the density at which the atoms start to expand with larger lattice depths. Our results suggest a new type of self-trapping, where quantum correlations suppress tunneling even when there are no density gradients.
\end{abstract}

\maketitle		

Ultra-cold atomic gases trapped by light are well-characterized many-body quantum systems. Without the disorder that is common in condensed matter systems, theoretical analyses of cold gases in equilibrium can be extremely accurate, although high powered numerical techniques must sometimes be employed~\cite{Bloch2008r}. Cold gas experiments are also well-suited to studying out-of-equilibrium dynamics, like the evolution of many-body correlations, since experimental time-scales are relatively slow yet faster than typical relaxation and decoherence rates~\cite{Ottl2005,Bakr2010,Hung2010}.  Significant progress in the description of the out-of-equilibrium dynamics of 1D bosonic and fermionic systems has been achieved thanks to the existence of exact solutions~\cite{Polkovnikov2011} and the development of numerical methods such as time dependent density-matrix renormalization group (t-DMRG)~\cite{Schollwock2005} and time-evolving block decimation (TEBD)~\cite{Vidal2003} methods. Those approaches fail in 2D and higher dimensions, where most out-of-equilibrium calculations rely on mean field, phase space methods, or approximate analytical techniques which are generally restricted to the weak interaction regime \cite{Polkovnikov2010}. The experiment-theory comparisons presented in this paper illustrate the breakdown of these approximate approaches,  highlighting the need for more computationally tractable methods for dealing with intermediate coupling. The particular way the theory deviates from the experiment strongly suggests a qualitatively new non-equilibrium effect.

Macroscopic self-trapping in quantum degenerate Bose gases, where mean-field energy gradients suppress tunneling, presents an interesting set of non-equilibrium phenomena, with analogs in non-linear photon optics~\cite{Musslimani2004,Hartmann2008} and Josephson junction arrays~\cite{Martinoli2000}. Self-trapping has been studied theoretically in all dimensions~\cite{Milburn1997,Smerzi1997,Raghavan1999,Trombettoni2001,Morsch2002,Alexander2006,Xue2008,Hipolito2010,Wuster2011} and experimentally in double-well systems~\cite{Albiez2005} and in arrays of 2D pancake-like Bose-Einstein condensates (BECs) created by a deep 1D lattice potential~\cite{Anker2005}. These self-trapping experiments are in the weak coupling limit, where mean-field theory clearly applies. In this letter, we experimentally and theoretically investigate self-trapping behavior in an array of coupled quasi-1D tubes created by a 2D optical lattice.  Atoms freely expand along the axis of the tubes, so that their density decreases with time, until they eventually become too dilute for self-trapping and expand ballistically transverse to the tubes.  In contrast to previous self-trapping work, our quasi-1D gases are in the intermediate coupling regime~\cite{Kinoshita2005}. To provide a baseline theoretical description of our experiments, we build a mean-field model of expanding coupled 1D gases.  We then incorporate fluctuations into our mean-field treatment via an approximation to the so called Truncated Wigner Approximation (TWA)~\cite{Blakie2008, Polkovnikov2010} by introducing random tube-to-tube phase fluctuations with a tunable magnitude at the initial time of the evolution.  At low lattice depths, small phase fluctuations improve the agreement with the experiment, while the dynamics in deeper lattices are better-described when the phases are maximally randomized. Still, the self-trapping seen in the experiment is more widespread across the array and persists to lower densities than mean-field based models predict.

\begin{figure}[htb]
\includegraphics[width=3.2in]{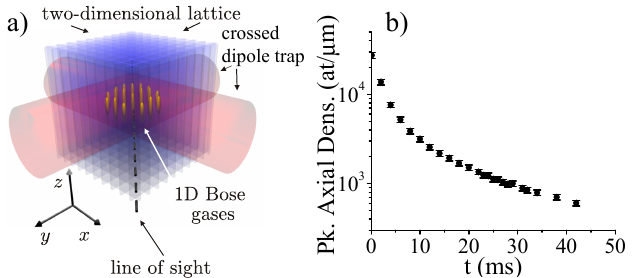}
\caption{(Color online) (a) Experimental setup (not to scale):  Atom clouds are confined in a crossed-dipole trap and partitioned by a blue-detuned optical lattice into a coupled array of vertical tubes.  (b) Peak transversely integrated axial density as a function of time for $V_0=13E_R$. The linear density drops rapidly with time as the atoms expand axially, mostly driven initially by their mutual interaction energy.} \label{fig1}
\end{figure}

It is instructive to first describe the pure mean field theory evolution qualitatively. When the BEC is loaded into the 2D lattice, the atoms are distributed among tubes with an approximately Thomas-Fermi profile. The tube-to-tube mean-field energy difference grows with the transverse density gradient. For a given peak density and tunneling energy, there is a critical distance along the lattice directions at which the mean-field imbalance between adjacent tubes is too large for tunneling to conserve energy. Atoms at larger radii will not tunnel radially outward, while atoms at smaller radii will start to expand, at some point reflecting from the self-trapped edge. A similar thing happens in 1D self-trapping~\cite{Anker2005}, but there are more features in the 2D generalization. Atoms at 45$^o$ to the lattice axes are only self-trapped at larger radii, since their density gradients are smaller in the lattice directions. The distributions therefore develop a density depression in the middle with a roughly square fort-like barrier around the edges. Since all atoms tunnel in some lattice direction, the self-trapped edges evolve with time. In the
absence of axial trapping, densities, and hence tube-to-tube density gradients, drop. Self-trapped edges are lost and the cloud ultimately expands ballistically in the $(x,y)$ plane.

Details of the experimental setup are given in \cite{Kinoshita2005a}, and the experimental geometry is depicted in Fig. \ref{fig1}a.  A BEC with a barely detectable impurity fraction and $N \approx 3.5 \times10^5$ $\mathrm{^{87}Rb}$ atoms in the $|F=1,m_F=1\rangle$ state is produced in a crossed optical dipole trap, created by the intersection of two 1.06 $\mu$m wavelength horizontal laser beams.  The trapping frequencies experienced by the BEC are $\omega_{x,y}=2\pi \times 38$ Hz in the horizontal plane, and $\omega_z= 2\pi \times 94$ Hz vertically.  A vertical magnetic field gradient of $30.5$~G$/$cm levitates the atoms. A two-dimensional square optical lattice is created using two slightly different frequency pairs of linearly polarized retro-reflected beams with $\mathrm{1/e^2}$ widths of 650 and 705~$\mathrm{\mu m}$, blue-detuned by 5.2~THz from the D2 transition in $^{87}$Rb. The 2D lattice is ramped up in time according to $s(t)=1/(1 + \alpha t^2)$, where $\alpha=4.15$~ms is the time constant. Non-adiabaticity in the lattice turn-on, measured by turning off the lattice with the reverse ramp, adds less than 4 nK$\cdot k_B$ of energy, which is small compared to the 1D chemical potential (150-250 nK$\cdot k_B$).

After the lattice is turned on, the crossed dipole trap beams are suddenly turned off so that the atomic distribution evolves in the 2D lattice alone.  After a time $t$, we image the in-situ spatial distribution of the cloud using high-intensity absorption imaging~\cite{note1}.  This technique involves illuminating the atoms with a resonant probe beam with intensity $I \gg I_{\rm{sat}}$, where $I_{\rm{sat}}$ is the saturation intensity.  After the beam passes through the atomic cloud, it is blocked by a $400~\mu$m diameter dark spot in the Fourier plane of a one-to-one imaging system. Using Babinet's principle, one can show that the detected signal is proportional to the square of the atomic density distribution integrated along the line of sight~\cite{note1}.  For the highest densities there is some lensing of the absorption beam, which we correct for by observing the axial expansion in a $35E_R$ lattice, where there is no transverse expansion. At our highest density, this correction is 10$\%$ of the measured root-mean-square (RMS) width.

\begin{figure}[htb]
\includegraphics[width=3.4in]{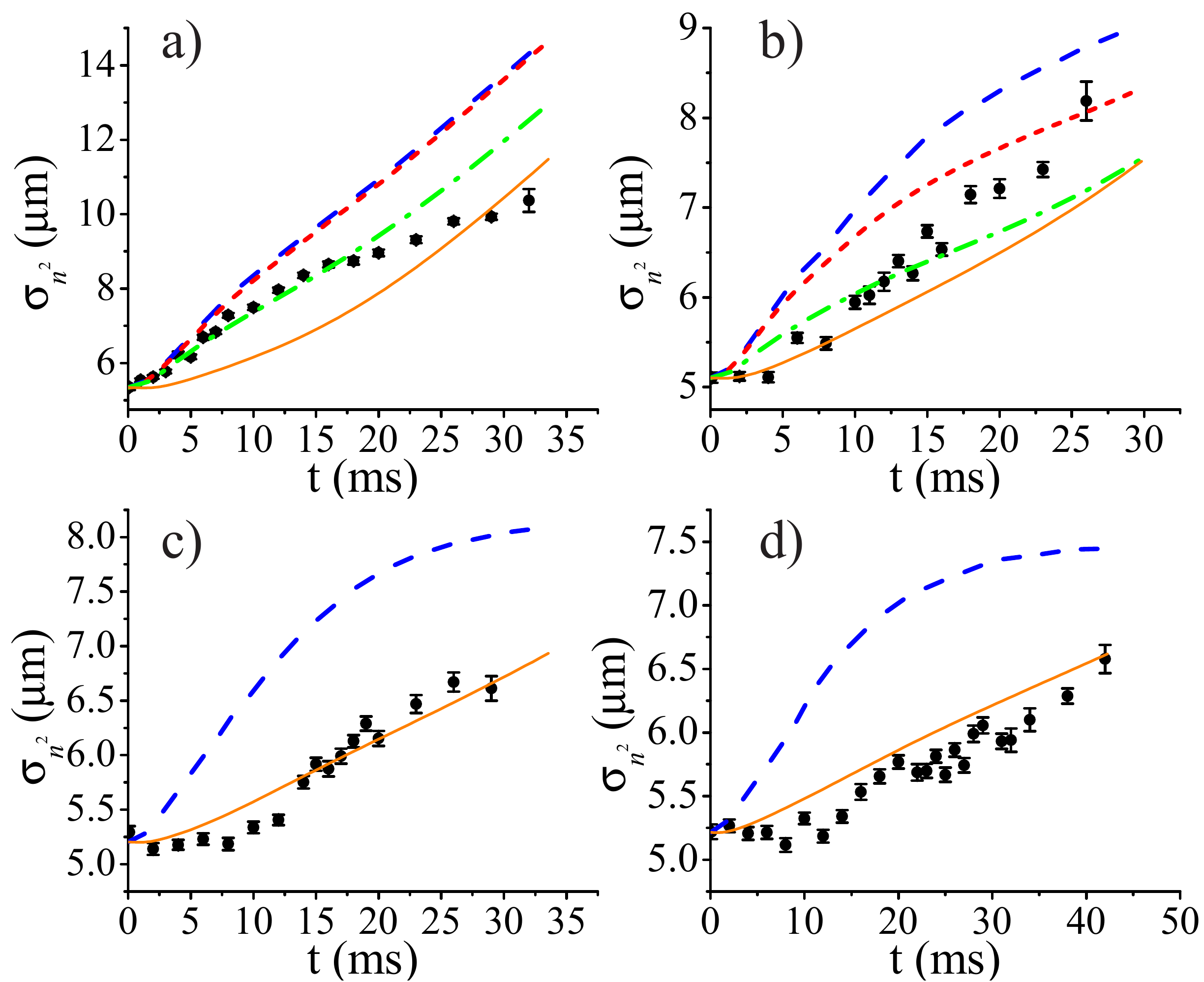}
\caption{ (Color online) Transverse RMS widths of the density distributions as a function of time for (a) $V_0=7.25~E_R$, (b) $9.25~E_R$, (c) $11~E_R$, (d) $13~E_R$.  Each data point is the average of 10 measurements. The error bars are the quadrature sum of the statistical noise and the uncertainty from the density dependence of the absorption beam lensing. There is an overall, systematic uncertainty associated with the imaging resolution of 0.5 $\mu$m; comparable shifts of the initial size simply shift the theoretical curves at early times (below ~10 ms), followed by a gradual
divergence of the curves at longer times. The blue dashed line is the result of a mean-field calculation of the dynamics with no random phase between the tubes ($\eta=0$).  The red dotted line is for $\eta=0.2$, the green dash-dotted lines are for $\eta=0.4$ (a) and $\eta=0.5$ (b), and the orange solid line is for $\eta=1$.} \label{fig2}
\end{figure}

We have taken images of the evolving atom distributions from along the $z$-axis, where the self-trapping depressions would be directly observed. We do not see them, although calculations show that our imaging system could resolve them if they were there. Instead, when viewed from along $z$ or transversely 45$^o$ from the lattice axes, the integrated density-squared profiles are featureless and well-fitted by Gaussians. All data displayed below is taken from the side view. We integrate the distributions over the central $\pm 10 \%$ of the vertical Thomas Fermi radius, so that the atoms we consider within each tube have approximately the same density.  We fit the resulting 1D distribution to a Gaussian and extract $\sigma_{n^2}$, the transverse RMS width.  In Fig.~\ref{fig1}b we plot an example of how the transversely-integrated axial peak density varies with time, which we derive from measuring the axial expansion of atoms. In Fig.~\ref{fig2}, we display $\sigma_{n^2}$ as a function of evolution time in the 2D lattice for lattice depths of $V_0=(7.25, 9.25, 11, 13) \, E_R$ (uncertainty $\pm 2.5 \%$), where $E_R=\hbar^2k^2/2M$ is the recoil energy and $M$ is the atomic mass.

The behavior of $\sigma{_n^2}$ is qualitatively different for $V_0=7.25~E_R$ compared to larger lattice depths. For $V_0=7.25~E_R$, the RMS width of the cloud increases quickly from $t=0$ and slows near $t=5$~ms, where the curve changes its concavity.  For $V_0=9.25, 11,$ and $13~E_R$, the RMS width remains constant to within experimental uncertainty for $t<t_c$, after which it increases linearly in time.  The time $t_c$ at which the cloud begins to expand increases as the lattice depth increases.  As noted earlier, the density decreases with time as the atoms expand axially in the tubes.  We can thus associate with the time $t_c$ a 3D density at the center of the cloud, $\rho(t_c)$.  We find that the ratio $\rho(t_c)/J = \{ 1190 \pm 310, 860 \pm 190, 910 \pm 210 \}$~$\mu m^{-3}E_R^{-1}$ for $V_0/E_R=\{9.25, 11, 13\}$, where $J$ is the tunneling matrix element.  The rough constancy of this ratio accords with the general concept of a self-trapping threshold~\cite{Trombettoni2001}, but the absence of an initial internal expansion shows that the pure mean-field description is incomplete.

The starting point for our mean-field theory is the Gross-Pitaevskii equation (GPE),
\begin{equation}
i\hbar\frac{\partial{\Psi}}{\partial{t}}=\left(-\frac{\hbar^2}{2M}\nabla^2+V(\vec{x})+g|\Psi|^2\right)\Psi \quad .
\end{equation}
Here $\Psi=\Psi(\vec{x},t)$ is the matter wave field, $g=\frac{4\pi a_{s} \hbar^{2}}{M}$, $ a_{s}$ is the s-wave scattering
length,  $V=V_0[\sin^2 ( \pi x /d)+\sin^2 ( \pi y /d)]$ is the  optical lattice potential, and $d$ is the lattice spacing. The interplay between free expansion along the tubes and the transverse lattice dynamics is accounted for self-consistently.

The wave field $\Psi(\vec{x},t)$ can be expanded in terms of Wannier functions along the optical lattice directions:

\begin{eqnarray}
\Psi(\vec{x},t)=\sum_{n,m}  W_{n,m}(x, y) \Phi_{nm}(z,t) \quad.
\label{ansatz}
\end{eqnarray}
$N_{n,m}(t)=\int dz |\Phi_{nm}(z,t)|^2$ is the number of atoms at the site $(n,m)$, satisfying the normalization condition $\sum_{n,m}N_{nm}=N$, with $N$ the total number of atoms.  $W_{n,m}(x, y)$ is the lowest band Wannier orbital localized at site $(n,m)$. We assume, justified by the experiment, that only the first-band Wannier orbitals are occupied.  We neglect any temporal dependence of the Wannier functions because the inter-well number/phase dynamics is much faster than the time associated with the change in shape of such functions.

Using the ansatz in Eq.~\ref{ansatz} and integrating within each tube over the lattice directions ($x$ and $y$), we get a set of coupled 1D GP equations:
\begin{eqnarray}
i\hbar \dot{\Phi}_{nm}&=&\left(-\frac{\hbar^{2}}{2M}\frac{\partial^2}{\partial^2 z} + U |\Phi_{nm}|^{2}\right)\Phi_{nm} \\
&-&J\left[\Phi_{n+1,m}+\Phi_{n-1,m}+\Phi_{n,m+1}+\Phi_{n,m-1}\right] \nonumber \label{dgp}
\end{eqnarray}
where $\Phi_{nm}=\Phi_{nm}(z,t)$, $J=\int d^2x W_{n,m}^{\ast }[\frac{\hbar ^{2}}{2M}\nabla^2- V] W_{n+1,m}$ is the nearest neighbor tunneling,  and $U =\frac{4\pi a_{s} \hbar^{2}}{d M}\int d^2x| W_{n,m}|^{4}$ is the onsite interaction energy. We numerically solve the GPEs by discretizing them along $z$ and calculating the kinetic energy term using a Fourier transform method. The widths obtained by assuming full adiabaticity during the lattice turn-on are larger than the experimentally observed widths. We therefore match the initial transverse width to the experimentally observed value, and adjust the initial axial size so that the energy released in the axial expansion matches the experimental value.

\begin{figure}[htb]
\includegraphics[width=3.5in]{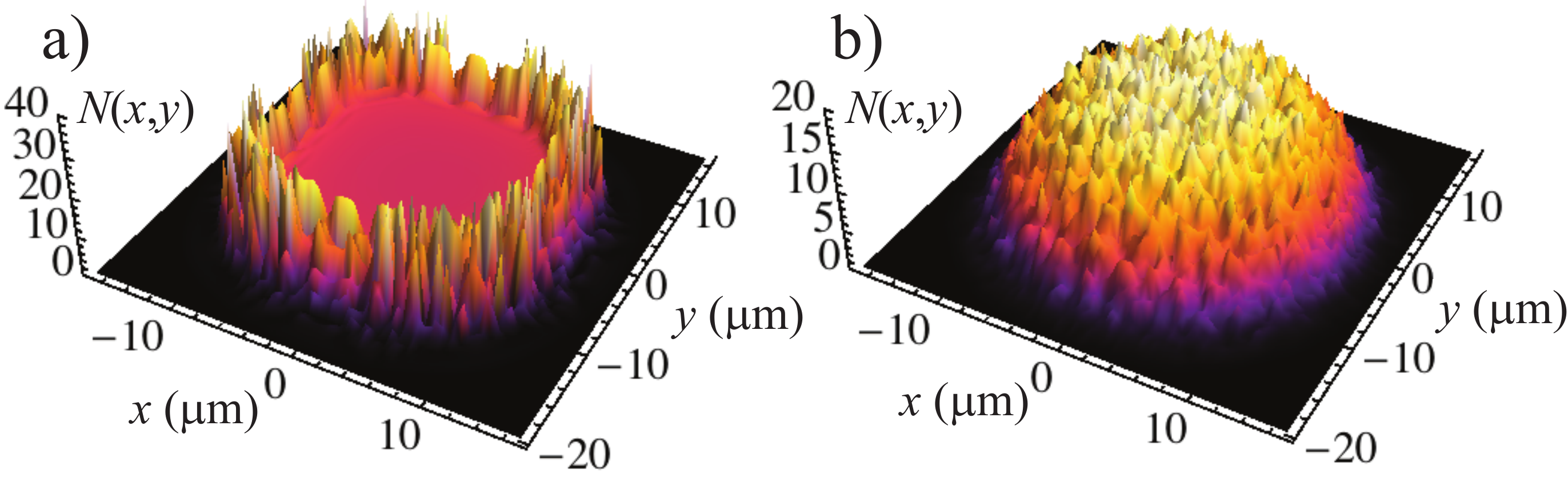}
\caption{ (Color online) Transverse density distributions 12.4~ms after release for $V_0=13E_R$. (a) $\eta$=0. (b) $\eta$=1. } \label{fig3}
\end{figure}

In the mean-field simulation~\cite{note4} the central part of the atom cloud starts to expand at $t=0$. Within several ms, self-trapping becomes evident with the formation of the previously mentioned roughly square fort-like structure (see Fig. \ref{fig3}a). Outwardly moving atoms cannot pass the steep density gradient, and they reflect back. All the while, the self-trapped edges evolve, leading to a complicated and ever-changing density distribution near the edges, and a slow spreading of the outer atoms. As the atoms expand axially along the tubes, the interaction energy decreases with time. Eventually the interaction energy differences between adjacent tubes become smaller than the tunneling energy, so that all self-trapping edges are lost and the cloud expands ballistically at a constant rate in the transverse direction.

In order to compare the simulation with the experimentally measured RMS widths, we extract widths from the simulated distributions in a way that mimics the measurement process (i.e., we square the density distribution and convolve it with a Gaussian of RMS width 1.7 $\mu$m, corresponding to the imaging system resolution). This process removes fine features from the theoretical curves that are not resolvable in the experiment, but it does not guarantee that the theoretical distributions are well-fitted by Gaussians. In fact, the pure mean-field theory calculations retain clear evidence of self-trapped squares even after this processing~\cite{note4}. The theoretical RMS widths are shown as blue dashed lines in Fig.~\ref{fig2}.  For all lattice depths presented, the agreement is poor.  For $V_0=7.25~E_R$, the mean-field calculation captures the qualitative dynamics, but overestimates the expansion rate at early times.  For $V_0=9.25,11$ and $13~E_R$, the calculation doesn't capture the qualitative dynamics. It greatly overestimates the initial expansion rates, and predicts that the RMS width increases much more slowly during the later ballistic expansion than is observed.

The mean-field treatment neglects thermal and quantum fluctuations, which can play large roles in the intermediate coupling limit. We note that it is hard to adiabatically transform a 3D gas into a 1D gas in theory~\cite{Polkovnikov2008} and, as described above, there is a small amount of irreversibility in turning on the lattice. Although the system is therefore not initially in its ground state, the observed self-trapping dynamics are insensitive to the details of the lattice turn-on and to the precise purity of the initial BEC, making quantum fluctuations the better candidate for the missing piece of the model. If we ignore tunneling and calculate the initial coupling strength, $\gamma$~\cite{Dunjko2001}, of atoms near where atoms are predicted by mean-field theory to have a self-trapped edge, it varies from 0.2 to 0.31 for $V_0=7.25~E_R$ and from 0.32 to 0.5 for $13~E_R$. Higher coupling implies larger phase variation along a tube. For small $V_0$, coherence among the tubes is maintained by tunneling. But as $V_0$ is increased and the tubes become more 1D, phase fluctuations along each tube will give rise to phase fluctuations from tube to tube.  This scenario is reminiscent of the deconfinement transition that occurs when there is an additional axial 1D lattice~\cite{Ho2004}, but the theory is less tractable out of equilibrium and with no 1D lattice, and as far as we know it has not been solved.

In an attempt to incorporate quantum fluctuation we use an approximate TWA, which gives quantum fluctuations to leading order when quantum dynamics are expanded around the classical (Gross-Pitaevski) limit. To implement the TWA we need to solve the mean field equations supplemented by random initial conditions distributed according to the Wigner transform of the initial density matrix.  Since the initial state is an interacting system, it is not simple to find the Wigner transform. Therefore, we accomplish an approximate TWA in an ad hoc way by introducing phase fluctuations into the initial conditions of our calculation by adding random phases among the tubes: $\phi_{nm}(t=0)\to 2 \pi \eta \Lambda$, where $\phi_{n,m}(0)$ is the initial phase of  the tube at lattice site $(n,m)$, $\Lambda $ is a uniformly distributed random variable between $0$ and $1$, and $\eta$ parameterizes the strength of the phase fluctuations. Within each tube the gas is still described by mean-field theory. When $\eta=0$, the initial conditions correspond to a fully coherent array of 1D gases and when $\eta=1$, we have a totally randomized, initially incoherent array.

As long as $\eta \gtrsim 0.4$, the squared and convolved theoretical distributions fit fairly well to Gaussians. We average the results from 20 randomized phase implementations, and display the RMS widths in Fig.~\ref{fig2} for various $\eta$ values. For $V_0=7.25~E_R$, $\eta \simeq 0.4$ best fits the early evolution, when there is self-trapping.  For larger $V_0$, $\eta=1$ clearly fits the early evolution best. Randomized phases cause initially random tunneling current directions throughout the tube array. Site-to-site density fluctuations rapidly develop, as seen in Fig.~\ref{fig3}b. Since there is no reflection from relatively sharp self-trapped edges, the acceleration of the cloud does not go negative (as it briefly does for $\eta$=0), but instead asymptotically approaches zero from above. Self-trapping is still lost when the density drops below a critical value that depends on $V_0$.

Although the modeled curves with the appropriate $\eta$ are never too far from the observed ones for $V_0=11E_R$ and $13E_R$, the experiment shows a more sudden and more delayed onset of ballistic expansion than the model, starting at a density that is at least three times smaller.  We have performed calculations with fluctuations in the initial atom number as well as phase in each tube and the curves are nearly unchanged.  Assuming a normal error distribution, the p-value is only 2$\times10^{-7}$ (6$\times10^{-19}$) that the points between 0 and 8 (12) ms in the $V_0=11E_R$ (13$E_R$) curves are consistent with mean field theory with maximal fluctuations. In contrast, the p-value is 0.45 (0.15) for them to lie on a horizontal line. The data suggests that the initial self-trapping is more complete than the model with maximal phase fluctuations predicts.

Detailed two-body correlations within each tube may be responsible. As the 2D lattice is turned on, correlations develop that make atoms avoid each other, at a kinetic energy cost, in order to avoid a larger mean-field energy cost. Since the wavefunction of a tunneling atom will not in general be appropriately correlated with the atoms in an adjacent tube, it would have to pay much of the mean-field energy cost that the correlations avoid. If that additional energy exceeds the tunneling energy, tunneling cannot proceed. This is a qualitatively new self-trapping phenomenon which, in the strong coupling limit, is similar to Pauli blocking.  It does not require density gradients, which is why the expansion of the central part of the tube bundle is fully suppressed until the overall density, and hence the mean-field cost of tunneling, gets sufficiently small. We speculate that such a process might tend to suppress mass transport whenever there is intermediate or strong coupling, regardless of the dimension~\cite{Fertig2005,Mun2007}.

In conclusion, we have measured the transverse expansion of quasi-1D arrays of atoms in the intermediate coupling regime.  We find coarse agreement with a mean-field model when we introduce intertube phase fluctuations to the initial conditions.  Modest phase fluctuations make the theory agree with experiment for $V_0=7.25~E_R$ for short times. For larger $V_0$, agreement with experiment is best with maximal initial phase fluctuations, which slow down the initial expansion. However, the experiment shows even less initial expansion, which suggests that there is strong self-trapping of a qualitatively different type, perhaps due to microscopic quantum correlations. This behavior could in principle be captured by a full TWA calculation; however, more work is needed to quantitatively model tunneling behavior in this quantum many-body regime.

{\em Acknowledgements.} The authors acknowledge fruitful discussions with A. Polkovnikov.  R.H. was supported by the AFOSR YI. D.S.W. acknowledges support from the NSF (PHY 11-02737), the ARO, and DARPA. A.M.R and S. Li acknowledge support from NSF, the AFOSR and the ARO (DARPA OLE ).


\end{document}